
\overfullrule=0pt
\magnification 1200
\baselineskip=24 true bp
\def\cp #1 #2 #3 {{\sl Chem.\ Phys.} {\bf #1}, #2 (#3)}
\def\jcp #1 #2 #3 {{\sl J.\ Chem.\ Phys.} {\bf #1}, #2 (#3)}
\def\jpa #1 #2 #3 {{\sl J. Phys.\ A} {\bf #1}, #2 (#3)}
\def\jsp #1 #2 #3 {{\sl J. Stat.\ Phys.} {\bf #1}, #2 (#3)}
\def\jtb #1 #2 #3 {{\sl J. Theor.\ Biol.} {\bf #1}, #2 (#3)}
\def\mb #1 #2 #3 {{\sl Molek.\ Biol.} {\bf #1}, #2 (#3)}
\def\jmb #1 #2 #3 {{\sl J.\ Mol.\ Biol.} {\bf #1}, #2 (#3)}
\def\mplb #1 #2 #3 {{\sl Mod.\ Phys.\ Lett.\ B} {\bf #1}, #2 (#3)}
\def\pmihas #1 #2 #3 {{\sl Publ.\ Math.\ Inst.\ Hung.\ Acad.\ Sci.}
{\bf #1}, #2 (#3)}
\def\pra #1 #2 #3 {{\sl Phys.\ Rev.\ A} {\bf #1}, #2 (#3)}
\def\pre #1 #2 #3 {{\sl Phys.\ Rev.\ E} {\bf #1}, #2 (#3)}
\def\prl #1 #2 #3 {{\sl Phys.\ Rev.\ Lett.} {\bf #1}, #2 (#3)}
\def\rmp #1 #2 #3 {{\sl Rev.\ Mod.\ Phys.} {\bf #1}, #2 (#3)}
\def\ltwid{\mathrel{\raise.3ex\hbox{$<$\kern-.75em\lower1ex\hbox{$\sim$}}}}
\def\eg{{\it e.\ g.}}\def\ie{{\it i.\ e.}}\def\etc{{\it etc.}}
\def\ref#1{${}^{#1}$}
\newcount\eqnum \eqnum=0  
\def\eqnoi{\global\advance\eqnum by 1\eqno(\the\eqnum)}
\def\eqnai{\global\advance\eqnum by 1\eqno(\the\eqnum {\rm a})}
\def\back#1{{\advance\eqnum by-#1 Eq.~(\the\eqnum)}}
\def\last{Eq.~(\the\eqnum)}                   

\newcount\refnum\refnum=0  
\def\refi{\smallskip\global\advance\refnum by 1\item{\the\refnum.}}

\newcount\rfignum\rfignum=0  
\def\rfigi{\medskip\global\advance\rfignum by 1\item{Figure \the\rfignum.}}

\newcount\fignum\fignum=0  
\def\figi{\global\advance\fignum by 1 Fig.~\the\fignum}
\def\pd#1#2{{\partial #1\over\partial #2}}      
\def\td#1#2{{d #1\over d #2}}      
\def\kp{k_+}
\def\km{k_-}
\def\k{\kp/\km}
\def\kg{\k\gg1}
\def\pxt{P(x,t)}
\def\pmt{P(m,t)}
\def\peq{P_{eq}(x)}
\def\pmeq{P_{eq}(m)}
\def\r{\rho}
\def\rt{\r(t)}
\def\req{\r_{eq}}
\def\rjam{\r_{\rm jam}}
\def\a{\alpha}
\def\b{\beta}
\def\l{\lambda}
\def\fpint{\int_0^{\infty}} 
\def\fplint{\int\limits_0^{\infty}} 
\def\logt{\log(t)}
\def\loglogt{\log\bigl(\logt\bigr)}

\centerline{\bf Collective Properties of Adsorption-Desorption Processes}
\bigskip
\centerline{\bf P.~L.~Krapivsky and E.~Ben-Naim}
\smallskip
\centerline{Center for Polymer Studies and Department of Physics}
\centerline{Boston University, Boston, MA 02215}
\vskip 1in
\centerline{ABSTRACT}
{\smallskip\noindent
A reversible adsorption-desorption parking process in one dimension
is studied. An exact solution for the equilibrium properties is obtained.
The coverage near saturation depends logarithmically
on the ratio between the adsorption rate, $\kp$, and the desorption rate,
$\km$, \hbox{$\req\cong 1-1/\log(\k)$}, when $\kg$.
A time dependent version of the reversible problem with
immediate adsorption ($\kp=\infty$) is also considered.
Both heuristic arguments and numerical simulations reveal a
logarithmically slow approach to the completely covered state,
\hbox{$1-\rt\sim 1/\logt$}.
}
{
\narrower\bigskip\noindent
}

\vfill\eject
\medskip\centerline{\bf I. Introduction}\smallskip

The adsorption of large particles such as colloids, proteins, latex spheres,
\etc\ on solid substrates is typically an irreversible process.\ref{1,2}~
Indeed, in a number of situations, the energetic barriers for desorption
are much higher than the corresponding barriers for adsorption.
Moreover, particles cannot adsorb on top of previously adsorbed ones.
This leads to the nonoverlapping irreversible random sequential
adsorption (RSA) models which have been studied intensively.
It was found that in arbitrary dimension, RSA processes reach a
jamming configuration, where further adsorption events are not possible.
The final coverage as well as the temporal approach to the jammed
state are of interest.\ref{2-8}~ Exact analytical results have been obtained
mainly in one dimension, where the problem is also known as the
``parking'' problem.\ref{2,7-8}~

It is clear that the usual RSA model provides an oversimplified description
of actual adsorption processes. A more realistic treatment should
incorporate various effects such as the transport properties of the particles,
the interaction between particles, and possible desorption from the
substrate to the bulk.\ref{9-13}~ Very recently, RSA models where
particles diffuse in the bulk and adsorb on the substrate were considered.
Interestingly, introduction of  bulk transport did not change the coverage
and the structure of the jammed configuration. However, it was found
that the approach to the jamming limit depends on the transport
properties of the particles.\ref{9,10}~

In this article, we study the influence of desorption on the one-dimensional
parking problem. Such a generalization is
appropriate for many physical, chemical and biological
systems.\ref{1-17}~ Allowing desorption makes the process manifestly
reversible and the system ultimately reaches an equilibrium state.
In the experimentally relevant desorption-controlled limit, the system
approaches the saturated state in a non-trivial manner.

The rest of this paper is organized as follows. In Section II, we
introduce the model, write the governing rate equations for the density
of empty intervals, and then find the exact steady state solution to
these equations. The primary result of this section is the weak
logarithmic dependence of the equilibrium coverage on the ratio of
the adsorption rate to the desorption rate, in the
desorption-controlled regime. In Section III, we describe the
temporal behavior of the system near saturation in the
desorption-controlled limit. To study the evolution in this limit,
we focus on a model with an infinite rate of adsorption
and a finite rate of desorption. A heuristic argument as well
as numerical  simulations show that the coverage slowly approaches
saturation, \hbox{$1-\rt\sim 1/\logt$}. Finally in Section IV, we
discuss our findings and further outlook.

\medskip\centerline{\bf II. The reversible parking problem}\smallskip

In the irreversible parking problem, identical particles park on a line
with an  adsorption rate $\kp$. Particles attempt to park with an
equal rate everywhere and a parking attempt fails if the space is
partially occupied by a previously adsorbed particle. We are interested
in the more general situation where particles are also allowed to
desorb with a desorption rate $\km$.
Particles desorb regardless of their local environment.
This system ultimately reaches
an equilibrium state independent of the initial conditions.
The primary aim of this study is to describe this final state and the
asymptotic approach towards it. To this end we will apply
the empty interval distribution method.\ref{14}~
This technique is often applied to RSA problems,
since simple and closed equations emerge. We generalize these equations
to describe the reversible case and solve the static equations.

Denote the density of empty intervals of  size exactly equal to $x$
at time $t$ by $\pxt$. Each empty interval borders a particle
to its left and to its right. Since in the adsorption-desorption
process one interval corresponds to one particle and since the total
density of particles and intervals is equal to unity one has
$$
1=\fplint dx\,(x+1)\pxt .\eqnoi
$$
Without loss of generality the size of particles is taken as the
unit of length. Moreover, from the same one-to-one mapping between
particles and intervals, the density of particles can be obtained
from the distribution function of empty intervals as
$$
\rt=\fplint dx \pxt. \eqnoi
$$

To write the evolution equations one has to account for
all possible processes leading to a change of $\pxt$. For $x<1$, an
interval disappears when either one of its neighboring particles
desorbs. On the other hand, an interval of size $x$ is
created when a particle parks at one of two specific locations on
an interval of size $y$ with $y>x+1$. These two processes give rise
to the following integro-differential equation for $\pxt$
$$
\pd{\pxt}{t}=-2\km\pxt+2\kp\int\limits_{x+1}^{\infty}dyP(y,t),
\qquad\qquad\qquad\qquad\qquad x<1.\eqnai
$$
Due to adsorption, intervals with length larger than a particle
size are destroyed with a rate $\kp(x-1)$.
In addition, two neighboring intervals can create a larger interval when the
intervening particle desorbs. However, the mere knowledge of the interval
distribution function is not sufficient for writing the rate equations.
We introduce the interval-interval distribution
function $P(y,z,t)$, defined as the density of pairs of
neighboring intervals of length $y$ and $z$, which are
separated by a single particle.
We further assume that this interval-interval distribution function is
proportional to the product of single interval densities. This
assumption, also known as the independence principle, has proved to be exact
in a number of RSA problems.\ref{18,19}~  In equilibrium at least,
one expects such a relation to be exact and it is a natural
starting point for investigating the time dependent problem.
With this assumption, one finds for the rate equation for $x>1$
$$
\eqalign{
\pd{\pxt}{t}=&-2\km\pxt+2\kp\int\limits_{x+1}^{\infty}dyP(y,t)\cr
&+{\km\over\rt}\int\limits_{0}^{x-1}dyP(y,t)P(x\!-\!1\!-\!y,t)
-\kp(x-1)\pxt \cr}
\qquad x>1. \eqno(3{\rm b})
$$
The convolution term involves the probability of finding a $y$ interval,
$P(y,t)$, and the normalized probability for its neighbor
to be of size $x-1-y$, \ie, \hbox{$P(x\!-\!1\!-\!y)/\fpint dx\pxt$}.

To verify that these equations satisfy the normalization condition
of \back2\ , one can check by direct integration of \last\ that
\hbox{${\partial\over\partial t}\fpint dx\,(x+1)\pxt=0$}.
Another useful check of self-consistency of the rate equations is
provided by integration of the rate equations over all lengths. This
gives the equation describing the change of the density,
$$
\td \rt t=-\km\rt+\kp\int\limits_1^\infty dx\,(x-1)\pxt, \eqnoi
$$
which can also be derived directly from the definition of the gap
distribution. Equation (4) is a generalization of the Langmuir mean-field
equation, $d\r/dt=-\km\r+\kp(1-\r)$, which is recovered by
setting the integral on the right-hand side  equal to $1-\rt$.

The steady-state interval distribution has to satisfy the static version
of \back1, that is, both sides of the equation vanish. Interestingly, the
simplest attempt, trying the Poissonian distribution
$\peq=\b\exp(-\a x)$, is successful. We emphasize that despite the two
different equations for $x<1$ and $x>1$, the solution is smooth at $x=1$.
{}From the normalization condition of Eq.~(1) the prefactor is determined
to be $\b=\a^2/(1+\a)$, while from \back2\ the value of $\a$ is found via a
transcendental equation involving $\k$. One can now write the exact
steady-state solution as
$$
\peq={\a^2\over 1+\a}\exp{(-\a x)} \qquad {\rm with}
\qquad\a\exp(\a)=\k. \eqnoi
$$
To obtain the density of particles in the steady-state we use the
correspondence between particles and intervals expressed in \back3\ .
Hence, it is easily found that $\req=\a/(1+\a)$. In the
desorption-controlled regime, $\k\gg 1$, one finds
$\a\simeq\log(\k)$, and consequently,
$$
\req\cong 1-{1 \over \log(\kp/\km)} \qquad \kg. \eqnoi
$$

Notice that in the limiting case $\k\to\infty$ the line is
completely filled with particles at the steady state, in contrast
to the case of no desorption where asymptotic coverage is
$\rho_{\rm jam}\cong 0.7475$.\ref{7-8}~
However, as the adsorption rate increases while the desorption
rate is kept fixed, the coverage increases very weakly,
since the corresponding correction is logarithmic.
Moreover, the gap distribution is regular at the origin, contrary to
the logarithmic divergence occurring in the irreversible case.
Thus, the presence of desorption, even if slight, significantly changes
the long-time behavior. Note also that in the adsorption-limited case,
the coverage displays an obvious linear dependence
on the rate ratio, $\req\cong\a\cong\k$.

To test the theoretical predictions we performed Monte-Carlo
simulations of the reversible parking process.
We have found
that the equilibrium properties are essentially identical to
the analytical results shown in Fig.~(1). We also confirmed
the Poissonian nature of the density of empty intervals.
Furthermore, the relaxation to the the steady state have been
investigated. As expected, the approach to the steady state coverage is
exponential  $\req-\rt\propto\exp(-t/\tau)$, where the relaxation time
$\tau$ appears to be proportional to $\k$ with possible logarithmic
corrections. Note that the  corresponding approach to the jammed
state in the irreversible case has a power-law dependence on time
$\rjam-\rt\propto t^{-1}$.
The distribution function $\pxt$ exhibits a transient
discontinuous derivative at $x=1$ due to the discontinuous structure of
\back3\ . That feature does not permit us to construct an analytic
solution for the time dependent problem.

We now outline the solution for the lattice
version\ref{15,16} of the reversible parking problem.
In this model, objects occupy $r$ lattice sites and will be
referred to as $r$-mers. Analogous to the continuous case,
$r$-mers land uniformly on a lattice with a rate $\kp$ and adsorb if all
sites are empty. We define $\pmt$ as the density of empty intervals of
exact length $m$. The equivalent to Eqs.~(1-3) simply follows
and we merely quote the final results. The density of empty intervals is
again Poissonian
$$
\pmeq={(1-\l)^2\over\l+r(1-\l)}\,\l^m\qquad {\rm with}
\qquad{\l^r\over1-\l}={\km\over\kp}.\eqnoi
$$

Using the equivalent of Eq.~(2) $\req=r\sum_0^{\infty} \pmeq$,
we can find the steady state coverage
\hbox{$\req=r(1-\l)/\bigl(\l+r(1-\l)\bigr)$}.
In the desorption-controlled case, $\kp/\km\gg1$ we find
that \hbox{$\req=1-(\km/\kp)^{1/r}/r$}. As the size of the $r$-mer
increases, the exponent of the power-law $r^{-1}$ decreases
and ultimately the aforementioned logarithmic nature is reached.
Indeed, the solution to the continuous case given by Eqs.~(5-6) can be found
from \last\ by taking the proper
limits $x\to m/r$ and $\kp\to\kp/r$.

\medskip\centerline{\bf III. Dynamics in the desorption-controlled
limit}
\smallskip

The second part of our study focuses on the dynamical properties
of the system in the desorption-controlled limit.
To attain this regime, immediate adsorption ($\kp=\infty$) is imposed
while the desorption rate is kept finite. Redefining the time
$t\to\km t$, corresponds to taking the desorption rate equal to unity.
The evolution may be divided into two stages.
First, the system instantaneously reaches a jammed state
and then desorption comes into play. Since both the final coverage
and the asymptotic behavior do not depend on the initial
conditions produced at the end of first stage, one can use any
initially jammed configuration satisfying the normalization
constraint of \back6. In the simulations, we have chosen a distribution
where all  gaps between particles are equal to $x_0$ with $x_0<1$,
$$
P(x,t=0)={\delta(x-x_0)\over 1+x_0}.\eqnoi
$$

Once a desorption event has occurred, either one or two adsorption
events are possible depending on the length of the two intervals
bordering the particle. We have adopted a natural scheme where the
first particle lands on a randomly picked segment in the open interval and
then, if feasible, the second particle lands randomly on the
remaining interval. Thus, after each desorption and subsequent
adsorption event(s) the number of particles is either left unchanged
or increased by one. Numerical simulations indicate
that this process approaches complete coverage in the long-time limit.

A simple heuristic argument explains the behavior of the system near
saturation. Let us imagine that a segment of length $L=N+2$ is
occupied by $N$ particles. Consider a typical situation where the gaps
between the particles are comparable ($\sim 1/N$).
A successful density-increasing
adsorption event may take place only as a result of a number of
ordered cooperative desorption-adsorption events. First, the
leftmost particle has to desorb and then adsorb near the left end
of the segment. Second, the next leftmost particle has to desorb and
then adsorb near the right side of the previous particle \etc .
Finally, a gap larger than the unit length is cleared
at the right edge of the segment, and an additional particle successfully
adsorbs.
The probability for the first event is $R/N$ since the leftmost particle
is desorbed first among $N$ particles; the probability for
the second event is $R/(N-1)$ \etc.
The factor $R<1$ accounts for the probability that the corresponding
adsorption event happens in the proper location. The total probability
for the cooperative event can be written as $p\sim R^N/N!$.
We can now evaluate the time dependence of $N(t)$.
The time required for a unit change in $N$ is inversely proportional to
$p$ and one has
$$
\td N t \propto {\Delta N\over\Delta t} \propto {p} \propto
\left({eR\over N}\right)^N.
\eqnoi
$$
In the last step the Stirling's formula $N! \sim (N/e)^N$ was used.
Solving \last\ yields the following asymptotic
relation for $N(t)$, \hbox{$N\sim \logt/\loglogt$}.
Since the uncovered fraction obeys $1-\rt\propto 1/N$,
an unexpectedly slow long-time behavior of the density emerges,
$$
1-\rt\propto {\loglogt\over\logt}.\eqnoi
$$

We describe now a simpler heuristic argument that also
predicts inverse logarithmic behavior but without double logarithmic
correction. We observe that when the system approaches to the completely
covered state the time interval between successive density-increasing
adsorption events increases. Hence, the system has time to
``equilibrate'' and one can approximate the gap distribution by
the Poissonian equilibrium distribution
\hbox{$\pxt=\a^2\exp(-\a x)/(1+a)$}.
To solve for the density we write \back6\ with a vanishing
desorption term
$$
\td \rt t=\int\limits_1^\infty dx\,(x-1)\pxt=
{1\over 1+\a}\exp{(-\a)}. \eqnoi
$$

On the other hand, one has $\rt=\a/(1+\a)$ from \back9\ and
consequently, the time derivative of the density can be expressed as
$d\rt/dt=(d\a/dt)/(1+\a)^2$. By equating the two expressions obtained
for $d\rt/dt$ we have
$$
\td \a t =(1+\a)\exp(-\a).
\eqnoi
$$
Solving this differential equation we arrive at the asymptotic
behavior of the density, that is, $\a\propto\logt$ and
$$
1-\rt\propto{1\over\logt}.\eqnoi
$$

Both estimates predict that the desorption-limited process
gives rise to a very slow inverse logarithmic approach to the completely
covered state. Thus we conclude that in the desorption-controlled
limit the dynamics of the system exhibits ``glassy'' relaxation.

Numerical simulations of the desorption-limited process were performed
using the following simple procedure.
A list of intervals $\{l_i<1\}$ is kept, while the
locations of the particles are ignored. A simulation step consists of
choosing randomly a pair of neighboring intervals, $\{l_i,l_{i+1}\}$.
Then the total length $l_i+l_{i+1}$ is redivided randomly to two
new intervals, $\tilde l_i$ and $\tilde l_{i+1}$. If one of these new
intervals is larger than unity, an additional particle adsorbs. Given
$\tilde l_i>1$, two new intervals are created randomly with their total
length equal to $\tilde l_i-1$. Time is updated after each event by the
inverse of the total number of intervals in the system. To verify the
predicted logarithmic approach to the saturated state, we write
$f(t)=\bigl(1-\rt\bigr)\logt$. The simulation results for $f(t)$ and
$f(t)/\loglogt$ are shown in Fig.~2(a). Both functions are slowly
varying in time. Since the former is an increasing function and the latter
a decreasing one, we conclude that the estimates (10) and (13) provide
the upper and lower bounds for the uncovered fraction, respectively.
It seems that the upper bound provides a slightly better approximation
for $f(t)$.

Similar to the general reversible case, the gap distribution is an
important characteristic of the process. Rather cumbersome rate
equations describe the time evolution of $\pxt$ in this case.
We do not write these rate equations since we have not been able to
obtain meaningful new results by analytical means.
Instead, in Fig.~2(b) we present
Monte-Carlo simulational results for the gap distribution function.
In the long-time limit, the gap distribution function appears to be
again Poissonian, at least over a significant range of the gap size.
We believe that this supports the argument leading
to the lower bound (13). However, the tail of the distribution,
crucial for the adsorption of new particles, cannot be determined
from these data.

The success of the Poissonian approximation suggests that the process is
mean-field in nature. To check this feature, a mean-field variant
of the above Monte-Carlo simulation was also considered. In this
model, the two randomly chosen intervals are not required to be neighbors.
In fact, the pair is chosen randomly from all available pairs.
Simulations have shown little quantitative change and practically
no qualitative change in the data similar to those presented on Fig.~2.
We performed other numerical experiments including, \eg, computation
of the pair correlation function. Again, simulational results
revealed an excellent agreement between the one-dimensional
and the mean-field versions.

\medskip\centerline{\bf IV. Discussion}
\smallskip

We have considered near saturation properties of two one-dimensional
adsorption-desorption processes. In the reversible case,
we have obtained an exact solution that exhibits a slowly varying
dependence of the coverage versus the rate ratio $\k$, when $\kp\gg\km$.
In the case of immediate adsorption, when the system approaches to the
completely covered state, we have shown that \hbox{$1-\rt\sim1/\logt$}.
We have performed numerical simulations both for the 1D model and for
the mean-field version of the model. Comparison of numerical
results have revealed a remarkably good agreement between both models.
Despite the apparent difference in the definition of the reversible model
and the desorption-controlled model the underlying physical mechanism
for increase in coverage is similar.
In a configuration where  the coverage is large, $\r\ltwid
1$, a growing number of cooperative desorption-adsorption events are
necessary for an additional adsorption to occur. Moreover, the slow
nature of the process allows for perfect mixing of different gaps
and hence leads to a Poissonian distribution of gaps.

The situation found in this adsorption-desorption process is
reminiscent of that
encountered in a number of one-dimensional systems where phase
transitions occur at zero temperature. Indeed, noting
that in the parking problem the rate ratio $\k$ plays the role
of temperature we conclude that all basic features of the reversible
model such as disordered steady state and exponential approach towards
equilibrium correspond to typical behaviors above the point of phase
transition. In the desorption-controlled case, the system reaches
the perfectly ordered final state while the approach towards it shows a
critical slowing down. Furthermore, this analogy suggests that in
the two-dimensional case a phase transition in
the adsorption-desorption system may take place at a finite $\k$.
The desorption-controlled limit in two dimensions seems very
interesting since the system can reach numerous metastable states
which include, \eg, two ordered perfect crystal structures,
triangular and square, and a number of polycrystalline structures
with a network of defects. Metastable states should be responsible
for remanance effects, slow relaxation, and sensitivity to initial
conditions. Elucidating properties of these metastable states and
their basins of attraction, ``glassy'' phase transitions, \etc,
is left for future studies.

\medskip\centerline{\bf Acknowledgments}\smallskip

After completing this article we became aware of similar results,
derived by independent means, for the dynamics in the
desorption-controlled limit.\ref{20}~
We thank J.~Talbot for letting us know of his work
and for a useful correspondance.
We are also thankful to D.~ben-Avraham
for pointing out some relevant references and especially to S.~Redner
for numerous discussions and for reading the manuscript.
We gratefully acknowledge ARO grant \#DAAH04-93-G-0021, NSF grant
\#DMR-9219845, and to the Donors of The Petroleum Research
Fund, administered by the American Chemical Society, for partial support
of this research.

\vfill\eject
\medskip\centerline{\bf References}\smallskip

\refi G.~Y.~Onoda and E.~G.~Liniger, \pra 33 715 1986 .
\refi for a comprehensive review see J.~W.~Evans, \rmp xx xxxx 1993 .
\refi J.~Feder, \jtb 87 237 1980 .
\refi Y.~Pomeau, \jpa 13 L193 1980 .
\refi R.~H.~Swedsen, \pra 24 504 1981 .
\refi E.~L.~Hinrichsen, J.~Feder, and T.~Jossang, \jsp 44 793 1986 .
\refi A.~R\'enyi, \pmihas 3 109 1958 .
\refi J.~J.~Gonzalez, P.~C.~Hemmer, and J.~S.~Hoye, \cp 3 228 1974 .
\refi P.~Schaaf, A.~Johner, and J.~Talbot, \prl 66 1603 1991 ;
B.~Senger, J.~-C.~Voegel, P.~Schaaf, A.~Johner,
A.Schmitt, and J.~Talbot, \pra 44 6926 1991 .
\refi G.~Tarjus and P.~Viot, \prl 68 2354 1992 .
\refi P.~Nielaba and V.~Privman, \mplb 9 533 1992 ;
      V.~Privman and M.~Barma, \jcp 97 6714 1992 .
\refi G.~Tarjus, P.~Schaaf, and J.~Talbot, \jcp 93 8352 1990 .
\refi R.~B.~Stinchcombe,  M.~D.~Grynberg, and M.~Barma, \pre 47 4018 1993 .
\refi E.~R.~Cohen and H.~Reiss, \jcp 38 680 1963 .
\refi A.~S.~Zasedatelev, G.~V.~Gurskii, and M.~V.~Volkenshtein, \mb 5
245 1971 .
\refi J.~D.~McGhee and P.~H.~von Hippel \jmb 86 469 1974 .
\refi T.~L.~Hill, {\it Cooperativity Theory in Biochemistry},
Springer-Verlag, New York (1985).
\refi J.~J.~Gonzalez and P.~C.~Hemmer,  \jcp 67 2469 1977 ;
J.~J.~Gonzalez and P.~C.~Hemmer,  \jcp 67 2509 1977 .
\refi J.~W.~Evans, D.~K.~Hoffmann, and D.~R.~Burgess, \jcp 80 936 1984 .
\refi X.~Jin, G.~Tarjus, and J.~Talbot, preprint.

\vfill\eject
\medskip\centerline{\bf Figure Caption}\smallskip

\rfigi The exact steady state coverage for the reversible parking
problem. $\req(\k)$ is plotted vs. $\km/\kp$ for $\kp>\km$
(solid line) and vs. $\k$ for $\km>\kp$ (dotted line).

\rfigi  Monte-Carlo simulation results for the infinite adsorption case.
The simulation was performed on a ring of length 100,000.
(a) The temporal approach to the fully occupied state. Shown are
$f(t)$, $f(t)=\bigl(1-\rt\bigr)\logt$ vs. $t$ (circles) and
$f(t)/\loglogt$ vs. (t) (squares).
(b) The gap distribution at the 65536 time step.

\vfill\eject\bye